\documentclass[runningheads,fleqn]{cl2emult}

\usepackage{makeidx}  
\usepackage{graphicx} 
\usepackage{subeqnar} 
\usepackage{multicol} 
\usepackage{cropmark} 
\usepackage{phys}     
\makeindex            

\begin{document}
\title*{Screening of a Point Charge:\protect\newline
        A Fixed-Node Diffusion Monte Carlo Study}
\toctitle{Screening of a Point Charge:\protect\newline
          A Fixed-Node Diffusion Monte Carlo Study}
\titlerunning{Screening of a Point Charge:
              A Fixed-Node Diffusion Monte Carlo Study}

\author{Erik Koch\inst{1} \and Olle Gunnarsson\inst{1}
        \and Richard M.~Martin\inst{2}}
\authorrunning{Erik Koch et al.}
\institute{Max-Planck-Institut f\"ur Festk\"orperforschung, 70569 Stuttgart, 
           Germany \and
           Department of Physics and Materials Research Laboratory,\\ 
           University of Illinois, Urbana IL 61801, USA}

\maketitle

\begin{abstract}
We study the static screening in a Hubbard-like model using fixed-node
diffusion Monte Carlo. We find that the random phase approximation is
surprisingly accurate even for metallic systems close to the Mott transition. 
As a specific application we discuss the implications of the efficient 
screening for the superconductivity in the doped Fullerenes. In the Monte 
Carlo calculations we use trial functions with two Gutzwiller-type parameters. 
To deal with such trial functions, we introduce a method for efficiently 
optimizing the Gutzwiller parameters, both in variational and in fixed-node 
diffusion Monte Carlo. 
\end{abstract}

\section{Introduction}

The random phase approximation (RPA) is widely used throughout solid state
physics. It properly describes the screening when the kinetic energy is much 
larger than the Coulomb energy. In the strongly correlated limit, however,
it is qualitatively wrong. Little is known for the intermediate regime, where
kinetic and Coulomb energy are comparable, and perturbative methods fail since
there is no small parameter. In such a situation quantum Monte Carlo is the
method of choice. Our goal is to investigate the screening in a strongly 
correlated system close to the Mott transition. 
To be specific we focus on the doped Fullerenes, like K$_3$C$_{60}$. 
These materials are strongly correlated and close to a Mott transition 
\cite{c60mott}, but they are also superconductors with quite large 
transition temperatures ($T_c\approx30\,\mathrm{K}$ for Rb$_3$C$_{60}$)
\cite{rmp}. The superconductivity is driven by the coupling to intramolecular 
phonons. These phonons mediate an effective electron--electron attraction 
which is, however, counteracted by the electron--electron repulsion, which is
large in such strongly correlated systems. Unlike conventional superconductors,
in the doped Fullerenes this repulsion is not reduced much by retardation 
effects.
Therefore efficient screening is important for reducing the electron--electron 
repulsion sufficiently to allow for an electron--phonon driven superconductivity
\cite{mu,screening}. Using quantum Monte Carlo we find that even for quite
strong correlations the RPA gives a surprisingly accurate description of 
the static screening on the metallic side of a Mott transition until the 
system is close to the transition \cite{screening}. Besides the immediate
consequences this result has for our understanding of the superconductivity 
in the doped Fullerides, it should also have quite general implications for the
physics of systems close to a Mott transition.

We start by introducing the model Hamiltonian used to describe the doped
Fullerenes. We briefly discuss the L\"owdin downfolding technique that
gives a general prescription for constructing low-energy Hamiltonians. 
Next we introduce the screening problem and discuss the results of
the Monte Carlo calculations. Finally we review the quantum Monte Carlo
methods used. We discuss the choice of the trial wave function and introduce
a very efficient method for optimizing the Gutzwiller-type parameters.

\section{Model Hamiltonian}

\begin{figure}[t]
 \includegraphics[width=.75\textwidth]{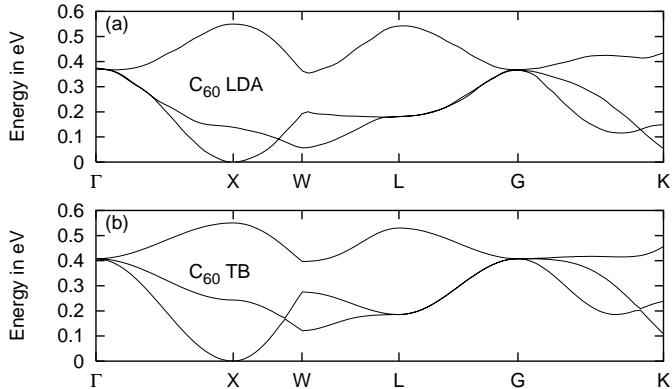}
 \caption[]{Band structure ($t_{1u}$ band) of solid C$_{60}$ (fcc)
            (a) as calculated ab initio using the
                local density approximation \cite{ldabands} and
            (b) using a tight-binding Hamiltonian with only $t_{1u}$
                orbitals \cite{TBparam}.}
 \label{c60bands}
\end{figure}

Fullerites\index{C$_{60}$} are crystals made of C$_{60}$ molecules 
on an fcc lattice. They are characterized by very weak inter-molecular
interactions. Therefore the discrete molecular levels merely broaden into 
narrow, well separated bands \cite{ldabands}. The valence band originates
from the lowest unoccupied molecular orbital, which is a 3-fold degenerate
$t_{1u}$ orbital\index{$t_{1u}$ orbital}. In doped Fullerenes, there are
alkali atoms sitting in the space between the C$_{60}$ molecules. They do 
not affect the band structure around the Fermi energy very much. Only
the filling of the $t_{1u}$ band changes, since each alkali atom donates 
its valence electron. Hence for K$_3$C$_{60}$ the (3-fold degenerate) 
$t_{1u}$ band is half-filled. We are mainly interested in the properties 
of these valence-band electrons. To simplify the description
we therefore want to get rid of the other bands. They can be
projected out by the L\"owdin downfolding\index{L\"owdin downfolding} 
technique \cite{lowdin}. The basic idea is to partition the Hilbert space 
into a subspace that contains the degrees of freedom that we are interested 
in (in our case the `$t_{1u}$-subspace') and the rest of the Hilbert space:
${\cal H}={\cal H}_0 \oplus {\cal H}_1$. We can then write the Hamiltonian 
of the system as
\begin{equation}
  H=\left(\begin{array}{cc}H_{00}&H_{01}\\ H_{10}&H_{11}\end{array}\right)\;,
\end{equation}
where $H_{ii}$ is the projection of the Hamiltonian onto subspace 
${\cal H}_i$, while the $H_{ij}$ $(i\ne j)$ contain the hybridization matrix 
elements between the two subspaces. Writing Green's function $G=(E-H)^{-1}$ 
in the same way, we can calculate the projection of $G$ onto ${\cal H}_0$ 
\cite{invpart}:
\begin{equation}
  G_{00}=\Big(E-\underbrace{[H_{00}+ 
                 H_{01}\,(E-H_{11})^{-1}H_{10}]}_{H_{\rm eff}(E)}\Big)^{-1} \; .
\end{equation}
We see that the physics of the full system is described by an effective
Hamiltonian $H_{\rm eff}(E)$ that operates on the subspace ${\cal H}_0$ only.
We have, however, to pay a price for this drastic simplification: the effective
Hamiltonian is energy dependent. In practice one approximates it with an
energy-independent Hamiltonian $H_{\rm eff}(E_0)$. This works well if we
are only interested in energies close to $E_0$.
In solid C$_{60}$ we have the fortunate situation that the bands
retain the character of the molecular orbitals, since the hybridization
matrix elements are small compared to the energy separations of the orbitals.
In fact we can neglect the other bands altogether and get the hopping matrix 
elements $t_{in,\,jn'}$ between the $t_{1u}$ orbitals $n$ and $n'$ on molecules
$i$ and $j$ directly from a tight-binding parameterization \cite{TBparam,A4C60}.
To demonstrate how well this works, Fig.\ \ref{c60bands} shows the comparison 
of the ab initio $t_{1u}$ band structure with the band structure 
obtained from the tight-binding \index{tight-binding} Hamiltonian with only
$t_{1u}$ orbitals.

A realistic description of the electrons in the $t_{1u}$ band also has
to include the correlation effects which come from the Coulomb repulsion
of electrons in $t_{1u}$ orbitals on the same molecule. The resulting
Hamiltonian which describes the interplay of the hopping of electrons 
and their Coulomb repulsion has the form
\begin{equation}\label{Hamil}
H=\sum_{\langle ij\rangle} \sum_{nn'\sigma} t_{in,jn'}\;
              c^\dagger_{in\sigma} c^{\phantom{\dagger}}_{jn'\sigma}
 +\;U\sum_i\hspace{-0.5ex} \sum_{(n\sigma)<(n'\sigma')}\hspace{-1ex}
       n_{i n\sigma} n_{i n'\sigma'} .
\end{equation}
The on-site Coulomb interaction\index{on-site Coulomb interaction} $U$ can 
be calculated within density functional theory \cite{calcU}. It is given by 
the increase in the energy of the $t_{1u}$ level per electron that is added 
to one molecule of the system. It is important to avoid double counting in 
the calculation of $U$. While the relaxation of the occupied orbitals and 
the polarization of neighboring molecules have to be included in the 
calculation, excitations within the $t_{1u}$ band must be excluded, since 
they are contained explicitly in the Hamiltonian (\ref{Hamil}).
The results are consistent with experimental estimates \cite{expU,lof}:
$U\approx 1.2-1.4\,\mathrm{eV}$. For comparison, the width of the $t_{1u}$ band
is in the range $W\approx 0.5-0.85\;eV$.
To properly describe K$_3$C$_{60}$ the effect of the orientational disorder
\cite{TBparam,Mazin} of the C$_{60}$ molecules in the crystal are built into 
the hopping matrix elements $t_{in,jn'}$. Multiplet effects are not included, 
since they tend to be counteracted by the Jahn-Teller effect, which is also 
neglected. 

In K$_3$C$_{60}$ the system has three electrons per molecule. In the limit 
of weak correlations ($U=0$), this corresponds to a metal with a half-filled
conduction band. In the atomic limit ($U\to\infty$) the Coulomb energy 
dominates, forcing every molecule to be occupied by exactly three electrons,
and suppressing any hopping. This is a Mott insulator. We therefore
expect a metal-insulator transition for some finite value of the Coulomb 
interaction $U$. For the model Hamiltonian (\ref{Hamil}) with parameters 
describing K$_3$C$_{60}$ it occurs for $U\approx 1.5-1.75\,\mathrm{eV}$ 
\cite{c60mott}. Given the estimates for the true value of $U$, K$_3$C$_{60}$
is therefore close to the Mott transition, in a correlated metallic state.

\section{Screening of a Point Charge}

We now investigate how efficient the screening is in a strongly correlated 
system like K$_3$C$_{60}$. To be specific we analyze how a test charge 
$q$ sitting on one molecule is screened by the conduction electrons in the 
$t_{1u}$ band. To describe the influence of the test charge situated on the
molecule at site $c$ we include an
additional term 
\begin{equation}\label{Hq}
  H_q = q\,U\;\sum_{n\sigma} n_{cn\sigma}
\end{equation} 
in the Hamiltonian (\ref{Hamil}). Determining the electron density at site $c$
for the system without test charge and for the system with a finite $q$ we 
find the screening 
\begin{equation}
  {\Delta n\over q}= {n_c(0)-n_c(q) \over q}\;.
\end{equation} 

\begin{figure}[bt]
\rotatebox{270}{\includegraphics[width=.58\textwidth]{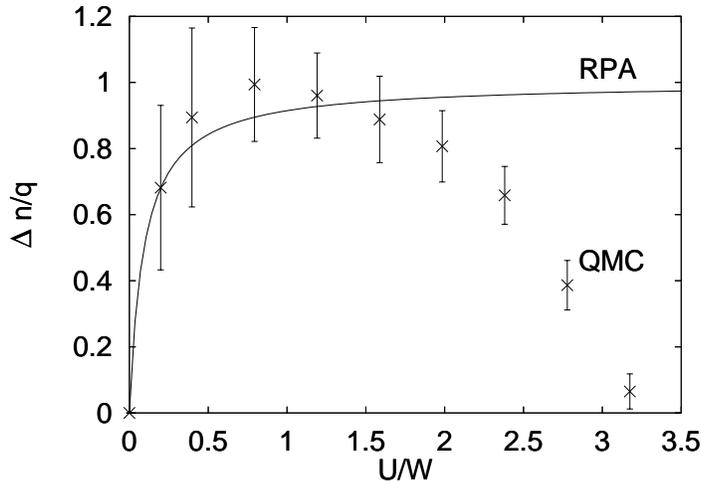}}
\caption[]{
 Screening charge $\Delta n$ on the site of the test charge 
 ($q=0.25\,e$) as a function of $U/W$, extrapolated to infinite 
 cluster size. The full curve shows the screening charge in the 
 RPA, obtained from Hartree calculations for the Hamiltonian 
 (\ref{scrhamil}). The crosses with errorbars give the results of 
 the QMC calculations.  The RPA screening remains rather accurate 
 up to $U/W\sim 2$, but fails badly for larger values of $U/W$. 
 The screening is very efficient for $U/W\sim 0.5-2.0$.  }
 \label{extrapolate}
\end{figure}

Let's first discuss the screening in the RPA. In the random phase 
approximation\index{random phase approximation} it only costs kinetic energy 
to screen the test charge. In the limit where a typical Coulomb integral $U$ 
is large compared with the band width $W$, the kinetic energy cost of 
screening is relatively small compared with the potential energy gain. 
Therefore, within the random phase approximation, the screening is very
efficient for large $U$. This means that as the test charge $q$ is introduced, 
almost the same amount of electronic charge moves away from the site: 
$\Delta n \approx q$ for large $U$ (see Fig.\ \ref{extrapolate}). The random 
phase approximation neglects, however, that when an electron leaves a site 
it has to find another site with a missing electron or there is an increase 
in Coulomb energy of the order of $U$. Thus the RPA is accurate for small 
values of $U/W$, while it is {\em qualitatively wrong} for large $U/W$. It 
is not clear what happens when Coulomb energy $U$ and band width $W$ are 
comparable.

To address this question we have performed quantum Monte Carlo calculations
for the combined Hamiltonian 
\begin{equation}\label{scrhamil}
H=\!\sum_{\langle ij\rangle} \sum_{nn'\sigma} t_{in,jn'}
              c^\dagger_{in\sigma} c^{\phantom{\dagger}}_{jn'\sigma}
 +U\sum_i\hspace{-1.0ex} \sum_{(n\sigma)<(n'\sigma')}\hspace{-2ex}
       n_{i n\sigma} n_{i n'\sigma'} 
 +q U \sum_{n\sigma} n_{cn\sigma}
\end{equation}
Since we are interested in the linear response, we should calculate the effect 
of an infinitesimally small test charge $q$. Because of the statistical error 
in a Monte Carlo calculation it is, however, difficult to determine the 
response to a small perturbation. To get a good signal-to-noise ratio, we 
would therefore like to use as large a test charge as possible. To estimate 
how large we can make $q$ and still be in the linear response regime, we have 
performed Lanczos calculations for a small system of 4 molecules, where exact
calculations are possible. Checking the response for different test charges
we find that for $q\leq 0.25\,e$ the response is practically linear. 
The quantum Monte Carlo calculations were then performed for large clusters of 
$N_\mathrm{mol}=$32, 48, 64, 72, and 108 molecules, corresponding to systems
with 96, 144, 192, 216, and 324 electrons, and $q=0.25$. To extrapolate the 
results $\Delta n(N_\mathrm{mol})$ to infinite systems size we used a 
finite-size 
scaling of the form $\Delta n=\Delta n(N_{\mathrm{mol}})+\alpha/N_\mathrm{mol}$.
The results are shown in Fig.\ \ref{extrapolate}. For rather small values of 
$U/W$ ($\sim 0.5-1.0$), the RPA somewhat underestimates the screening. Such 
a behavior is also found in the electron gas \cite{Hedin}. For intermediate 
values of $U/W$ ($\sim 1.0-2.0$) the RPA gives surprisingly accurate results. 
For larger $U/W$ the screening rapidly breaks down, and the RPA becomes 
qualitatively wrong, as discussed above.

The efficient screening almost up to the Mott transition which occurs for
$U/W\sim 2.5$ \cite{c60mott} has profound implications for the 
superconductivity in the alkali doped Fullerenes. We will only give qualitative
arguments here, a more detailed discussion can be found in \cite{screening}. 
In BCS theory the superconducting transition temperature is given by
\begin{equation}\label{Tc}
  T_c\propto e^{-1/(\lambda-\mu^\ast)}\;,
\end{equation}
where $\lambda=N(0)\,V$ describes the electron--phonon coupling, which mediates 
the effective electron--electron attraction, $\mu^\ast=N(0)\,U^\ast$ is the 
Coulomb pseudopotential that describes the repulsive Coulomb interaction 
between electrons, and $N(0)$ is the density of states at the Fermi energy. 
Since the electrons couple to intramolecular phonons, the coupling constant
$V$ is to a good approximation a molecular property. Therefore increasing the
density of states $N(0)$ will increase the electron--phonon coupling $\lambda$.
This can be achieved by increasing the lattice constant of the solid 
Fullerene. Taking the C$_{60}$ molecules further apart will decrease the 
hopping matrix elements $t_{in,jn'}$ thus narrowing the $t_{1u}$ band and
correspondingly increasing the density of states. If also $U^\ast$ was 
a molecular quantity, i.e.\ independent of the lattice constant, and
if $V>U^\ast$ which, of course, is the prerequisite for superconductivity, 
then according to (\ref{Tc}) increasing the lattice constant would raise $T_c$. 
Such a variation of the transition temperature with the lattice constant is
indeed found experimentally \cite{rmp}. Here the variation of the lattice 
constant $a$ is achieved by applying hydrostatic pressure (to reduce $a$) 
or by using \lq larger' alkali metals like Rb and Cs to increase the lattice 
constant compared to K$_3$C$_{60}$ (chemical pressure). This suggests that
by inserting even more bulky ions or molecules, like NH$_3$, into the doped
Fullerenes, could be increase $T_c$ even further. As it turns out, however,
$T_c$ rapidly {\em decreases} with $a$ when the lattice constant becomes
too large, and eventually superconductivity disappears. This drop in $T_c$ 
can be understood as a natural consequence of the breakdown of the screening 
close to the Mott transition. In the doped Fullerenes $U^\ast$ is essentially 
given by $U\,(1-\Delta n/q)$, where $U$ is the unscreened Coulomb matrix 
element (cf.\ the interaction term in the Hamiltonian) and $\Delta n/q$ 
describes the screening by the $t_{1u}$ electrons. $U$ is practically 
independent of the lattice constant, while the screening efficiency changes 
with the band width as shown in Fig.\ \ref{extrapolate}. For $U/W$ in the 
region $1.0-2.0$ the screening is almost RPA-like and does not vary very much, 
which means that in this region $U^\ast$ is almost independent of the lattice 
constant $a$. In this region $T_c$ thus increases with $a$. For even smaller 
band width, or correspondingly larger lattice constants, the screening 
rapidly becomes inefficient. Now $U^\ast$ increases with the lattice
constant, leading to a decrease of $\lambda-\mu^\ast$. Hence for $a$ too large,
$T_c$ rapidly decreases with $a$ and eventually vanishes. Since the screening
only breaks down close to the metal-insulator transition, we have the 
interesting situation that {\em $T_c$ peaks close to a Mott transition!}

The efficient screening found in the Monte Carlo calculations also explains
why the coupling to the alkali-phonons is weak. Each C$_{60}$ molecule is
surrounded by 14 alkali ions with very weak force constants. When an electron
arrives on a molecule one would therefore expect that the surrounding
alkali ions respond strongly. This was, however, not confirmed by experiment
\cite{alkaliisotope}. That result can be naturally understood as an effect 
of the efficient screening: When an electron arrives on a molecule other
electrons leave, effectively leaving the molecule almost neutral. The
alkali ions then only see a small change in the net charge and therefore
respond weakly, leading to a small electron--phonon coupling. But being
molecular crystals, doped Fullerenes also have intramolecular phonons. 
Some of those intramolecular phonons shift the $t_{1u}$ levels in such
a way that the center of gravity of the energy levels is not changed.
These are the modes that are not screened by the transfer of charged.
They are therefore the modes that drive superconductivity in the Fullerides.

\section{Quantum Monte Carlo}

We now turn to the question of how the results shown in Fig.\ \ref{extrapolate}
were obtained. To keep the notation simple we will discuss the different
methods for a simple Hubbard model (only one orbital per site, next neighbor 
hopping matrix elements $t_{ij}=-t$):
\begin{equation}\label{Hubbard}
  H=-t\;\sum c^\dagger_i c_j + U\sum n_{i\uparrow} n_{i\downarrow} \; .
\end{equation}
The generalization to Hamiltonians like (\ref{Hamil}) is straightforward.

The first step in the quantum Monte Carlo approach is to identify a trial 
function $\Psi_T$. For the Hubbard model that function should balance the 
opposing tendencies of the hopping term and the interaction: 
Without interaction (i.e.\ for $U=0$) the ground state of the Hamiltonian 
(\ref{Hubbard}) is the Slater determinant $\Phi$ that maximizes the
kinetic energy. Without hopping ($t=0$) the interaction is minimized.
Since only doubly occupied sites, i.e.\ sites with $n_{i\uparrow}=1$
and $n_{i\downarrow}=1$, contribute to the Coulomb energy, 
the electrons are distributed as uniformly as possible over the lattice
to minimize the number of double occupancies. A good compromise between
these two extremes is to start from the non-interacting wavefunction $\Phi$
but reduce the weight of configurations $R$ with large double occupancies
$D(R)$. This leads (up to normalization) to the Gutzwiller 
wavefunction\index{Gutzwiller wavefunction} \cite{GWF}:
\begin{equation}\label{GWF}
 \Psi_T(R) =  g^{D(R)}\;\Phi(R) ,
\end{equation}
with $g\in(0,1]$ the Gutzwiller parameter. In configuration space the 
Coulomb term in the Hamiltonian is given by $U\,D(R)$. Thus 
the Gutzwiller factor reflects the interaction term in the Hubbard Hamiltonian.
In this spirit we can also construct trial functions for Hamiltonians with
additional terms like the screening Hamiltonian (\ref{scrhamil}).
We add a second Gutzwiller factor that reflects the interaction with the
test charge $q\,U\,n_c(R)$:
\begin{equation}\label{scrwf}
  \Psi_T(R) = g^{D(R)}\,h^{n_c(R)}\;\Phi(R) \;.
\end{equation}
Changing the additional Gutzwiller factor $h$ we can vary the occupation 
of the site with the test charge.

\subsection{Variational Monte Carlo}

Since the Gutzwiller factor, like the interaction term, is diagonal in
configuration space, we have to perform a sum over all configurations $R$
in order to calculate the energy expectation value for a Gutzwiller 
wavefunction:
\begin{equation}\label{Evar}
 E_T = {\langle\Psi_T|H|\Psi_T\rangle \over \langle\Psi_T|\Psi_T\rangle}
     = {\sum_R E_{\rm loc}(R)\;\Psi_T^2(R) \over \sum_R \Psi_T^2(R)} \;,
\end{equation}
where we have introduced the local energy for a configuration $R$
\begin{equation}\label{Eloc}
 E_{\rm loc}(R) 
 = \sum_{R'} {\langle\Psi_T|R'\rangle\,\langle R'|H|R\rangle
             \over \langle\Psi_T|R\rangle} 
 = \sum_{R'}\!'\;t\;{\Psi_T(R')\over\Psi_T(R)} + U\,D(R) \;.
\end{equation}
Since the number of configurations $R$ grows exponentially with system-size, 
the summations in (\ref{Evar}) can only be done explicitly for very small 
systems.  For larger problems we use variational Monte Carlo\index{variational 
Monte Carlo} \cite{VMC}. The idea is to perform a random walk in the space of 
configurations, with transition probabilities $p(R\to R')$ chosen such
that the configurations $R_\mathrm{VMC}$ in the random walk have the probability
distribution function $\Psi_T^2(R)$. Then
\begin{equation}\label{Evmc}
 E_\mathrm{VMC} = 
{\sum_{R_\mathrm{VMC}} E_{\rm loc}(R_\mathrm{VMC})\over\sum_{R_\mathrm{VMC}} 1}
 \approx
  {\sum_R E_{\rm loc}(R)\;\Psi_T^2(R) \over \sum_R \Psi_T^2(R)} 
 = E_T . 
\end{equation} 
The transition probabilities can be determined from detailed balance
\begin{equation}\label{detailedbalance}
  \Psi_T^2(R)\,p(R\to R') = \Psi_T^2(R')\,p(R'\to R)
\end{equation}
which is fulfilled by the choice
$p(R\to R')={1/N}\;\min[1,\Psi_T^2(R')/\Psi_T^2(R)]$, with
$N$ being the maximum number of possible transitions. 
It is sufficient to consider only transitions between configurations that
are connected by the Hamiltonian, i.e.\ transitions in which one electron 
hops to a neighboring site. The standard prescription is then to propose a 
transition $R\to R'$ with probability $1/N$ and accept it with probability 
$\min[1,\Psi_T^2(R')/\Psi_T^2(R)]$. This works well for $U$ not too large.
For strongly correlated systems, however, the random walk will stay for long
times in configurations with a small number of double occupancies $D(R)$, since
most of the proposed moves will increase $D$ and hence be rejected with
probability $\approx 1-g^{D(R')-D(R)}$.

There is, however, a way to integrate-out the time the walk stays in a 
given configuration. To see how, we first observe that for the local energy
(\ref{Eloc}) the ratio of the wavefunctions for all transitions induced by 
the Hamiltonian have to be calculated. This in turn means that we also
know all transition probabilities $p(R\to R')$. We can therefore eliminate
any rejection (i.e.\ accept with probability one) by proposing moves with
probabilities
\begin{equation}
  \tilde{p}(R\to R') = {p(R\to R')\over\sum_{R'} p(R\to R')} 
                     = {p(R\to R')\over 1-p_{\rm stay}(R)} \;.
\end{equation}
Checking detailed balance (\ref{detailedbalance}) we find that now we are
sampling configurations $\tilde{R}_\mathrm{VMC}$ from the probability 
distribution
function $\Psi_T^2(R)\,(1-p_{\rm stay}(R))$. To compensate for this we assign
a weight $w(R)=1/(1-p_{\rm stay}(R))$ to each configuration $R$. The energy 
expectation value is then given by
\begin{equation}
 E_T \approx 
 {\sum_{\tilde{R}_\mathrm{VMC}} w(\tilde{R}_\mathrm{VMC})\,
    E_{\rm loc}(\tilde{R}_\mathrm{VMC}) \over
  \sum_{\tilde{R}_\mathrm{VMC}} w(\tilde{R}_\mathrm{VMC})} \;.
\end{equation} 
The above method is quite efficient since it ensures that in every Monte Carlo
step a new configuration is created. Instead of staying in a configuration 
where $\Psi_T$ is large, this configuration is weighted with the expectation
value of the number of steps the simple Metropolis algorithm would stay there.
This is particularly convenient for simulations of systems with strong 
correlations: Instead of having to do longer and longer runs as $U$ is 
increased, the above method produces, for a fixed number of Monte Carlo 
steps, results with comparable error estimates.

\subsubsection*{Correlated sampling}

So far we have only specified the form of the trial function $\Psi_T$.
The goal of a variational calculation is now to identify the parameters that 
result in the best trial function. A criterion for a good trial function is 
e.g.\ a low variational energy. To find the wavefunction that minimizes the 
variational energy we could perform independent VMC calculations for a set 
of different trial functions. It is, however, difficult to compare the 
energies from these calculations since each VMC result comes with its own 
statistical errors. This problem can be avoided with correlated 
sampling\index{correlated sampling} \cite{corrsmpl}.
The idea is to use the same random walk for calculating the expectation value
of all the different trial functions. This reduces the {\em relative} errors 
and hence makes it easier to find the minimum.

Let us assume we have generated a random walk $R_\mathrm{VMC}$ using $\Psi_T$ as
the trial function. Using the same random walk, we can then estimate the energy
expectation value (\ref{Evmc}) for a different trial function $\tilde{\Psi}_T$,
by introducing the reweighting factors $\tilde{\Psi}_T^2(R)/\Psi_T^2(R)$: 
\begin{equation}\label{corrsmpl}
 \tilde{E}_T \approx 
  {\sum_{R_\mathrm{VMC}} \tilde{E}_{\rm loc}(R)\,\tilde{\Psi}_T^2(R)/\Psi_T^2(R)
   \over 
   \sum_{R_\mathrm{VMC}}                         \tilde{\Psi}_T^2(R)/\Psi_T^2(R)
  }\;.
\end{equation}
with
\begin{eqnarray}
  \tilde{E}_{\rm loc}(R)
  &=&\sum_{R'} t {\tilde{\Psi}_T(R')\over\tilde{\Psi}(R)} + U\;D(R) \\
  &=&\sum_{R'} t {\Psi_T(R')\over\Psi_T(R)}\;
       {\tilde{\Psi}_T(R')/\Psi_T(R') \over \tilde{\Psi}_T(R)/\Psi_T(R)}
     + U\;D(R) \nonumber
\end{eqnarray}

We notice that (also in $\tilde{E}_{\rm loc}$) the new trial function 
$\tilde{\Psi}_T$ appears only in ratios with the old $\Psi_T$. 
For trial functions (\ref{GWF}) that differ only in the Gutzwiller factor 
this means that the Slater determinants cancel, leaving only powers
$(\tilde{g}/g)^{D(R)}$. Since $D(R)$ is {\em integer} we can then rearrange 
the sums in (\ref{corrsmpl}) into polynomials in $\tilde{g}/g$. To find the
optimal Gutzwiller parameter we then pick a reasonable $g$, perform a VMC run
for $\Psi_T(g)$ during which we also estimate the coefficients for these 
polynomials. We can then calculate $E(\tilde{g})$ by simply evaluating the
ratio of the polynomials. Since there are typically only of the order of some
ten non-vanishing coefficients this method is very efficient. 

\begin{figure}[t]
  \rotatebox{270}{\includegraphics[width=.5\textwidth]{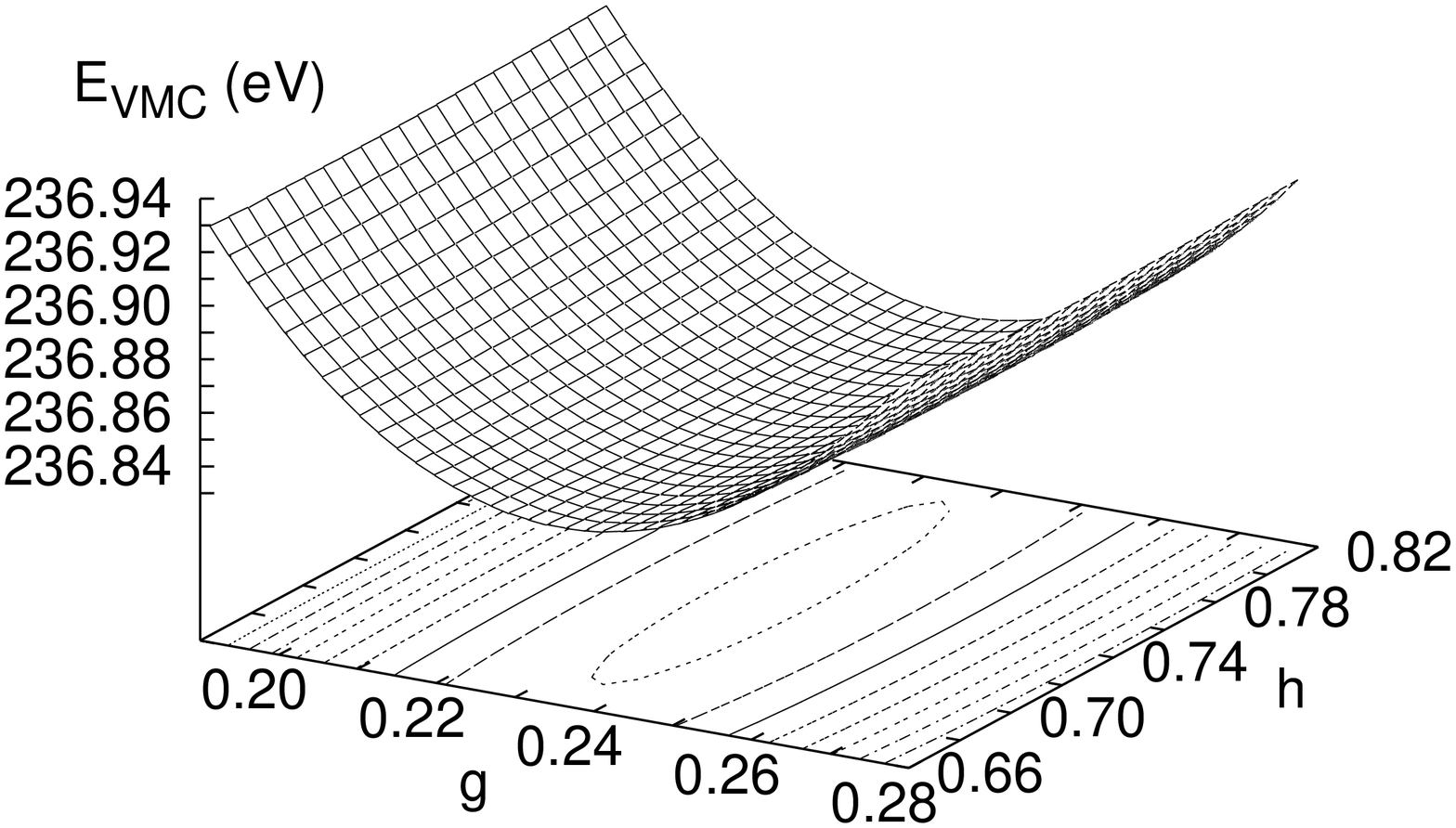}}
  \rotatebox{270}{\includegraphics[width=.5\textwidth]{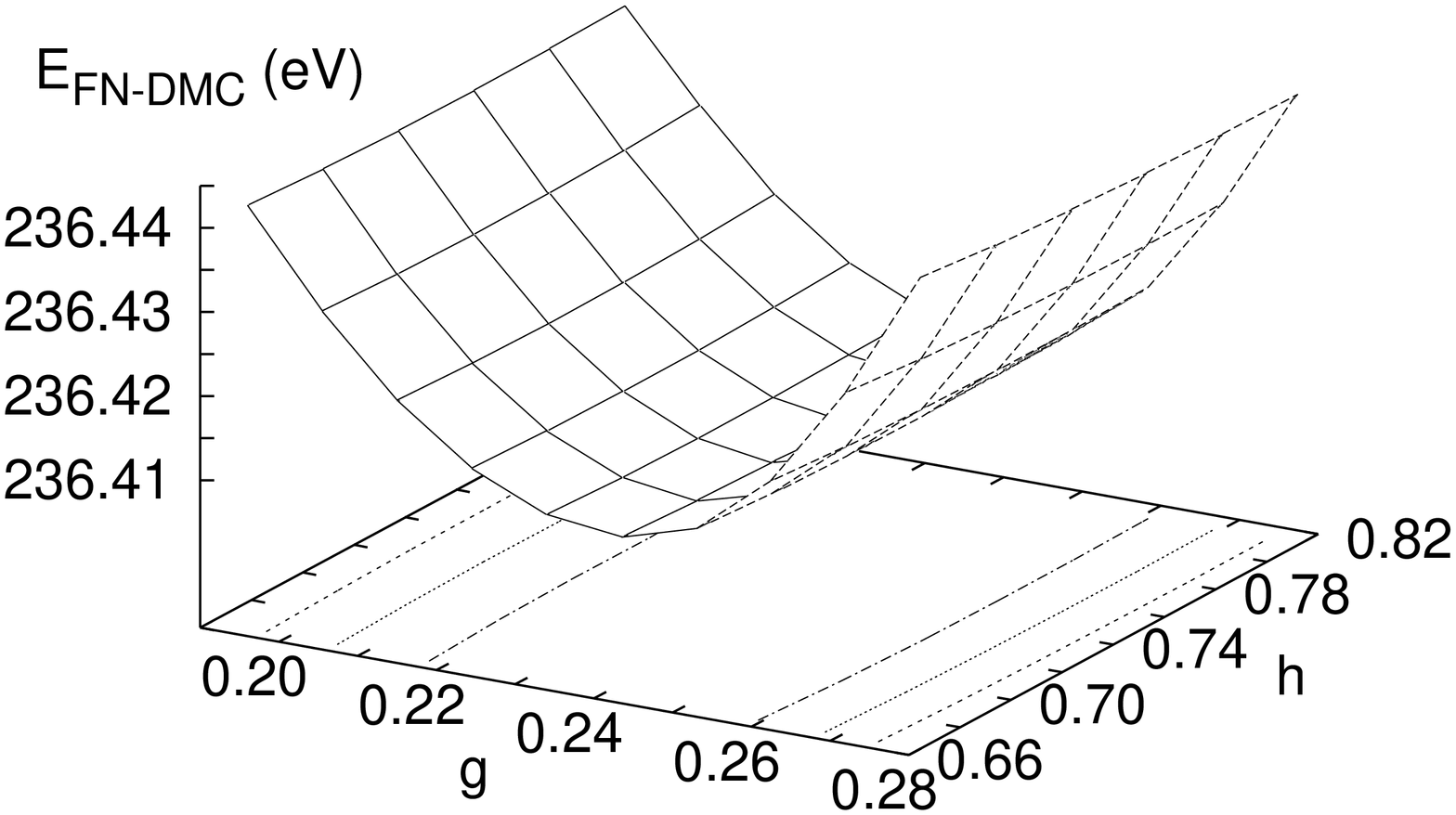}}
  \caption[]{\label{screen}
             Correlated sampling for the parameters $g$ and $h$ in
             the generalized Gutzwiller wavefunction
             $|\Psi_T\rangle= g^D h^{n_c}\,|\Phi\rangle$, cf.\ eqn.\ 
             (\ref{scrwf}), in variational (upper plot) and fixed-node
             diffusion Monte Carlo (lower plot). The plots show the energy
             as a function of the Gutzwiller parameters $g$ and $h$, both as
             surfaces and contours. The calculations were done for an fcc
             cluster of 64 molecules with $96+96$ electrons (half-filled
             $t_{1u}$-band), an on-site Hubbard interaction 
             $U=1.25\,\mathrm{eV}$, and a test charge of $q=1/4\,\mathrm{e}$.
            }
\end{figure}

The idea of rewriting the sum over configurations into a polynomial can be
easily generalized to trial functions with more correlation factors of 
the type $r^{c(R)}$, as long as the correlation function $c(R)$ is 
integer-valued on the space of configurations. As a specific example of
how the method works in practice, Fig.\ \ref{screen} shows the result of 
a correlated sampling run for the trial function (\ref{scrwf}).

\subsection{Fixed-node diffusion Monte Carlo}

Diffusion Monte Carlo\index{diffusion Monte Carlo} \cite{GFMC} allows us, 
in principle, to sample
the true ground state of a Hamiltonian. The basic idea is to use a projection
operator that has the lowest eigenstate as a fixed point. For a lattice problem
where the spectrum is bounded $E_n\in[E_0,E_{\rm max}]$, the projection is
given by
\begin{equation}\label{proj}
  |\Psi^{(n+1)}\rangle = [1-\tau(H-E_0)]\;|\Psi^{(n)}\rangle
    \;;\quad |\Psi^{(0)}\rangle=|\Psi_T\rangle .
\end{equation}
If $\tau<2/(E_{\rm max}-E_0)$ and $|\Psi_T\rangle$ has a non-vanishing overlap
with the ground state, the above iteration converges to $|\Psi_0\rangle$. There
is no time-step error involved. 
Because of the prohibitively large dimension of the many-body Hilbert space, 
the matrix-vector product in (\ref{proj}) cannot be done exactly. Instead, we
rewrite the equation in configuration space
\begin{equation}\label{iter}
  \sum |R'\rangle\langle R'|\Psi^{(n+1)}\rangle
        = \sum_{R,R'} |R'\rangle
          \underbrace{\langle R'|1-\tau(H-E_0)|R\rangle}_{=:F(R',R)}
          \langle R|\Psi^{(n)}\rangle
\end{equation}
and perform the propagation in a stochastic sense: $\Psi^{(n)}$ is
represented by an ensemble of configurations $R$ with weights $w(R)$.
The transition matrix element $F(R',R)$ is rewritten as a transition
probability $p(R\to R')$ times a normalization factor $m(R',R)$. The iteration 
(\ref{iter}) is then stochastically performed as follows: For each $R$ we pick 
a new configuration $R'$ with probability $p(R\to R')$ and multiply its weight 
by $m(R',R)$. Then the new ensemble of configurations $R'$ with their 
respective weights represents $\Psi^{(n+1)}$. Importance 
sampling\index{Importance sampling}
decisively improves the efficiency of this process by replacing $F(R',R)$ with 
$G(R',R)=\langle\Psi_T|R'\rangle\,F(R',R)/\langle R|\Psi_T\rangle$, so 
that transitions from configurations where the trial function is small
to configurations with large trial function are enhanced:
\begin{equation}
  \sum |R'\rangle\langle\Psi_T| R'\rangle\langle R'|\Psi^{(n+1)}\rangle
   = \sum_{R,R'} |R'\rangle\,G(R',R)\,
                 \langle\Psi_T|R\rangle\,\langle R|\Psi^{(n)}\rangle .
\end{equation}
Now the ensemble of configurations represents the product $\Psi_T\,\Psi^{(n)}$.
After a large number $n$ of iterations the ground state energy is then 
given by the mixed estimator\index{mixed estimator}
\begin{equation}\label{mixedest}
 E_0 = {\langle\Psi_T|H|\Psi^{(n)}\rangle \over \langle\Psi_T|\Psi^{(n)}\rangle}
      \approx {\sum_R E_{\rm loc}(R)\;w(R) \over \sum_R w(R)} .
\end{equation}
As long as the evolution operator has only non-negative matrix elements
$G(R',R)$, all weights $w(R)$ will be positive. If, however, $G$ has  
negative matrix elements there will be both configurations with positive and
negative weight. Their contributions to the estimator (\ref{mixedest})
tend to cancel so that eventually the statistical error dominates, rendering
the simulation useless. This is the infamous sign problem\index{sign problem}.
A straightforward way to get rid of the sign problem is to remove the
offending matrix elements from the Hamiltonian, thus defining a new Hamiltonian
$H_{\rm eff}$ by
\begin{equation}
   \langle R'|H_{\rm eff}| R\rangle = \left\{
     \begin{array}{cc}
                 0           & \mbox{ if $G(R',R)<0$} \\
      \langle R'|H| R\rangle & \mbox{ else} 
     \end{array}\right.
\end{equation}
For each off-diagonal element $\langle R'|H| R\rangle$ that has been removed,
a term is added to the diagonal:
\begin{equation}
  \langle R|H_{\rm eff}|R\rangle 
       = \langle R|H|R\rangle
       + \sum_{R'} \Psi_T(R')\langle R'|H|R\rangle/\Psi_T(R) .
\end{equation}
This is the fixed-node approximation\index{fixed-node approximation} for 
lattice Hamiltonians \cite{FNDMC}. $H_{\rm eff}$ is by construction free of the
sign problem and variational, i.e.\ $E_0^{\rm eff}\ge E_0$. The equality holds
if $\Psi_T(R')/\Psi_T(R)=\Psi_0(R')/\Psi_0(R)$ for all $R$, $R'$ for which 
$G(R',R)$ is negative. 

Fixed-node diffusion Monte Carlo for a lattice Hamiltonian thus means that
we choose a trial function from which we construct an effective Hamiltonian
and determine its ground state by diffusion Monte Carlo.
Because of the variational property, we want to pick the $\Psi_T$ such that
$E_0^{\rm eff}$ is minimized, i.e. we want to optimize the trial function, or,
equivalently, the effective Hamiltonian. Also here we can use the concept of 
correlated sampling. For optimizing the Gutzwiller parameter $g$ we can 
even exploit the idea of rewriting the correlated sampling sums into 
polynomials in $\tilde{g}/g$, that we already have introduced in VMC.
There is, however, a problem arising from the fact that the weight
of a given configuration $R^{(n)}$ in iteration $n$ is given by the product
$w(R^{(n)})=\prod_{i=1}^n m(R^{(i)},R^{(i-1)})$. Each individual normalization
factor $m(R',R)$ can be written as a finite polynomial, but the order of the
polynomial for $w(R^{(n)})$ increases steadily with the number of iterations.
It is therefore not practical to try to calculate the ever increasing number
of coefficients for the correlated sampling function $E^{(n)}(\tilde{g})$. 
But since we still can easily calculate the coefficients for the $m(R',R)$, 
we may use them to evaluate $E^{(n)}(\tilde{g})$ in each iteration on a set 
of predefined values $\tilde{g}_i$ of the Gutzwiller parameter. Figure 
\ref{screen} shows an example, again for the more general trial function 
(\ref{scrwf}).

\subsubsection{Extrapolated estimator}

So far we have only considered estimators for the total energy. For 
determining the screening, however, we need to know the charge density
$n_c$ on site $c$. It is no problem to calculate the expectation value
$n_c(\mathrm{VMC})=\langle\Psi_T|\hat{n}_c|\Psi_T\rangle$ in variational Monte
Carlo. Diffusion Monte Carlo gives, however, only the mixed estimator
$n_c(\mathrm{DMC})=\langle\Psi_T|\hat{n}_c|\Psi_0\rangle$. Since the density 
operator does not commute with the Hamiltonian, the mixed estimator is different
from expectation $n_c(\mathrm{QMC})=\langle\Psi_0|\hat{n}_c|\Psi_0\rangle$. 
For a good choice of the trial function the true expectation value can
be determined using the extrapolated estimator\index{extrapolated estimator}
$n_c(\mathrm{QMC})\approx 2\;n_c(\mathrm{DMC})-n_c(\mathrm{VMC})$.  

\begin{figure}[t]
 \rotatebox{270}{\includegraphics[width=.5\textwidth]{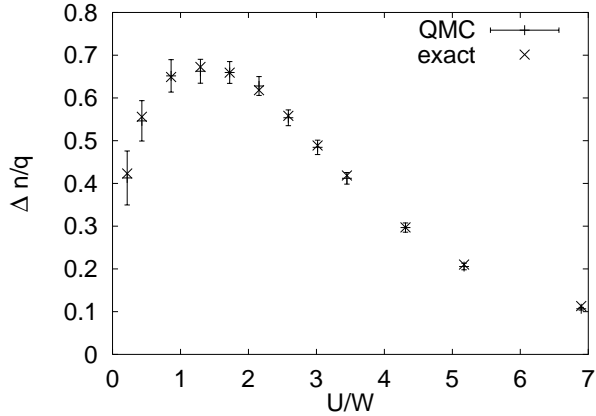}}
\caption[]{
 Screening charge $\Delta n$ on the site of the test charge 
 ($q=0.25\,\mathrm{e}$) as a function of $U/W$, where $U$ is the Coulomb 
 interaction and $W$ is the band width. Exact diagonalization 
 and QMC calculations have been performed for four molecules 
 (12 electrons). The figure shows that the QMC calculations 
 are quite accurate over the whole range of $U/W$.}
 \label{lcheck}
\end{figure}

To test the accuracy of this approach to the screening problem, which besides 
the extrapolated estimator also involves the fixed-node approximation, we have
compared the results of the quantum Monte Carlo calculations with the exact
results from exact diagonalization for a small system of four molecules
(12 electrons). As illustrated by Fig.\ \ref{lcheck}, the QMC calculations
for determining the screening charge are very accurate.

\subsubsection*{Acknowledgment.}
This work has been supported by the Alexander-von-Humboldt-Stiftung 
under the Feodor-Lynen-Program and the Max-Planck-Forschungspreis, 
and by grant No.\ DEFG02-96ER45439 from the Department of Energy.

\clearpage
\addcontentsline{toc}{section}{Index}
\flushbottom
\printindex

\end{document}